\documentclass[letter]{aa}  

\usepackage{graphicx}

\usepackage{txfonts}
\usepackage{subcaption}

\begin{document} 

   \title{Harmonic phase diagnostics of long secondary periods} 

   \subtitle{Testing predictions of oscillatory convective dipole modes in the OGLE sample}

   \author{B. Courtney-Barrer\inst{1,2,3}, X. Haubois\inst{1}, M. Ireland\inst{2}, P. Wood\inst{2}
          }

   \institute{European Southern Observatory, Alonso de Cordova 3107 Vitacura, 19001, Santiago, Chile\\
              \email{benjamin.courtney-barrer@mq.edu.au}
         \and
             Research School of Astronomy and Astrophysics, Australian National University, Canberra 2611, Australia
             \and
            Australian Astronomical Optics \& School of Mathematical and Physical Sciences, Macquarie University, NSW 2109, Australia
             }

   \date{Received 27 February 2026; Accepted 30 March 2026}

% \abstract{}{}{}{}{} 
% 5 {} token are mandatory
      \abstract
      {Long secondary periods (LSPs) in luminous red giants remain the only major class of long-period stellar variability without a secure physical origin. Competing hypotheses include binaries with dusty companions and oscillatory convective dipole modes.}
    % aims heading (mandatory)
    {We identify the physical and geometric conditions under which oscillatory convective dipole modes produce distinctive harmonic signatures that contrast with those expected from binary systems, and apply this diagnostic to a filtered subset of the OGLE-III LSP sample to identify examples consistent with oscillatory convective dipole modes.}
    % methods heading (mandatory)
    {We model the geometric flux modulation from oscillatory convective dipole modes and map the range of inclinations, temperature amplitudes, and observing wavelengths for which harmonic features are observable. Using OGLE-III I-band light curves, we require statistically significant power at both  sequence D and its harmonic, keeping a filtered sample of 249 stars (2.1\% of the ridge-selected sample). We apply iterative Lomb-Scargle and weighted Fourier decomposition to isolate the fundamental and harmonic components. The relative phase ($\Delta\phi$) between these distinguishes secondary maxima predicted by an inclined dipole from secondary minima caused by eclipsing or ellipsoidal binary systems.}
    % results heading (mandatory)
    {The majority of high amplitude stars in the filtered subset show $\Delta\phi$ consistent with secondary minima produced by binary systems. However, a small but statistically non-negligible subset exhibits $\Delta\phi$ consistent with secondary maxima that are difficult to reconcile by eclipsing or ellipsoidal binaries, and instead match the geometric predictions for highly inclined, non-rotating oscillatory convective dipole modes with temperature amplitudes consistent with published models.}
    %conclusion
    {}

   \maketitle

%
%________________________________________________________________
%/Users/bencb/Documents/long_secondary_periods/SECONDARY_ECLIPSE_THERMAL_MODE.py
%/Users/bencb/Documents/long_secondary_periods/why_galaxies_care_about_agb/fourier_fits_ogle.py
\section{Introduction}
Roughly one third of luminous red giants exhibit an enigmatic, long-timescale variability, commonly referred to as the long secondary period (LSP), superimposed on their primary pulsations and spanning several hundred to over a thousand days. Despite extensive study, LSPs remain the only class of large-amplitude stellar variability without a widely accepted physical explanation \citep{takayama_2020_LSP_thermal_dipole,pawlak_2021_connection_LSP_RGB_evolution, soszyski_2021_binary_origin_LSP}. In the period–luminosity (PL) diagram of long-period variables (LPVs), LSP stars define a distinct and well-populated ridge (sequence D) \citep{1999_wood_LPV_PL_classification}. In contrast to other sequences which are now understood as radial or non-radial pulsation modes \citep{trabucchi_2019_PL_puls_modes, nicholls_2009}, \citet{woods_2004_origin_of_LSP} showed that adiabatic pulsations cannot account for sequence D, and the physical origin of sequence D remains unresolved. Two observationally motivated hypotheses have emerged:
\begin{itemize}
\item Binary scenario: LSPs arise from a low-mass companion, often accompanied by dust clouds that cause periodic obscuration. \citep[e.g.][]{soszyski_2021_binary_origin_LSP, goldberg_2024_betelgeuse_lsp,decin_2025_binary}
\item Oscillatory convective mode scenario: LSPs correspond to dipole ($\ell=1$) highly nonadiabatic g- modes \citep[e.g.][]{saio_2015_oscillatory_convective_modes,takayama_2020_LSP_thermal_dipole}
\end{itemize}
Recent studies have highlighted the presence of secondary minima in LSP light curves, particularly at infrared wavelengths, as compelling evidence for the binary scenario \citep{soszyski_2021_binary_origin_LSP}. If these systems host brown dwarf companions that have accreted from a stellar wind, there is an additional challenge in reconciling the required angular momentum with the progenitor population of giant planets \citep{fulton_2021_planet_stat}. Most discussions to date have focused on finding a universal mechanism, with limited attention to the possibility of multiple LSP origins coexisting across the AGB and RGB populations.  In this Letter, we focus on a distinguishing observational feature: the presence of harmonics in sequence D stars. In binaries, these harmonics can arise from secondary eclipses, producing a phase-locked secondary minimum in the light curve as shown by \citet{soszyski_2021_binary_origin_LSP}. However, no similar mechanism has been previously proposed to explain such harmonic structure under the oscillatory convective mode hypothesis. We show here that oscillatory dipole modes can naturally produce strong harmonics through purely geometric effects linked to viewing inclination. Crucially, these harmonics manifest not as secondary minima, but as secondary maxima in the light curve, yielding a distinctive phase offset between the fundamental and its harmonic. This provides a new diagnostic capable of separating the binary and convective-mode scenarios. We identify and analyze examples of such signatures in OGLE light curves.

\section{Harmonics on sequence D}
\begin{figure}[h]
    \centering
    \includegraphics[width=8cm]{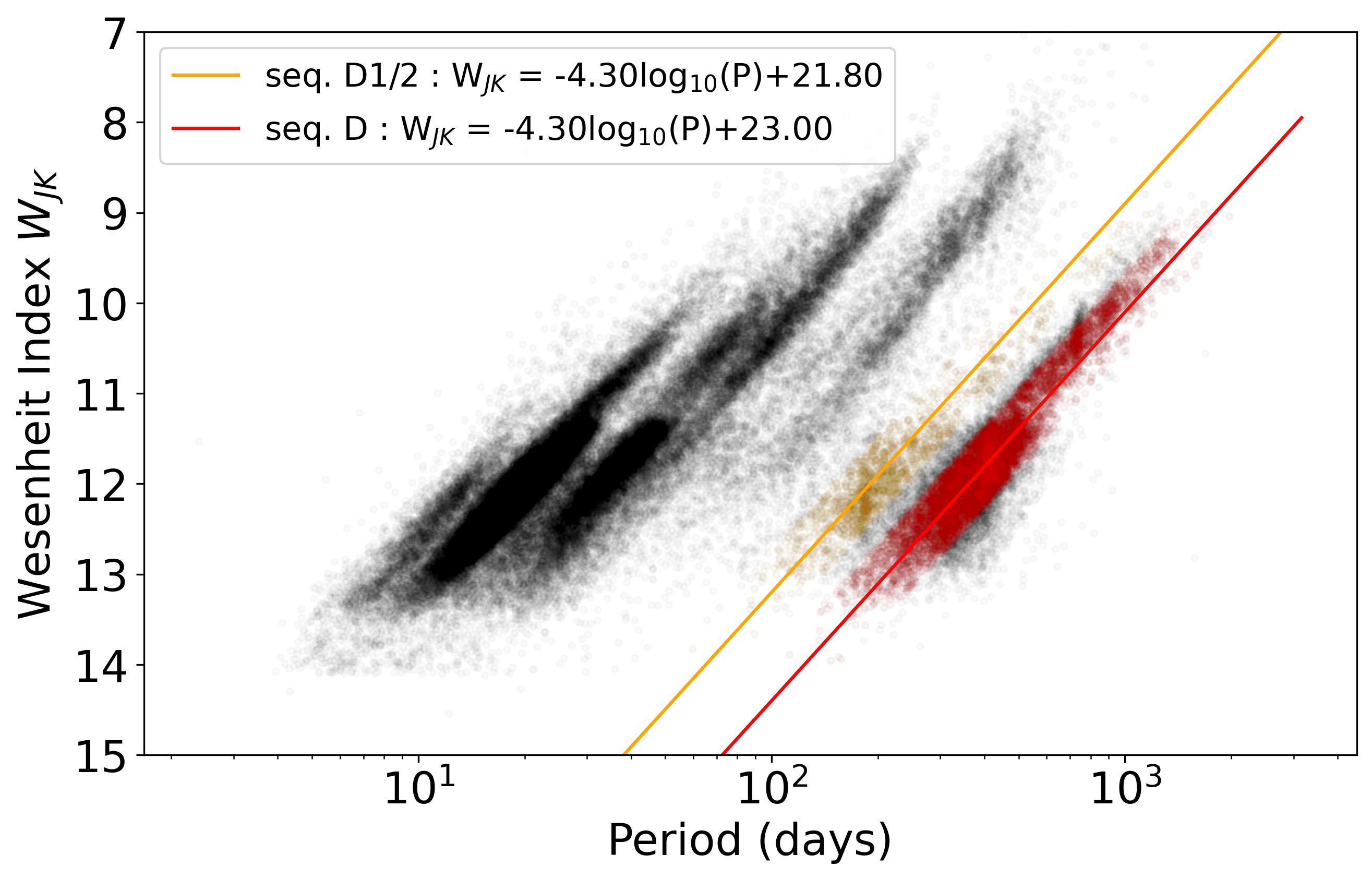}
    \caption{ Sequence D (red) in the PL diagram along with a sub-sequence at close to the second harmonic of sequence D which we highlight in yellow and label sequence D1/2. This ridge largely overlaps with the classical sequence E associated with ellipsoidal binaries, but is highlighted to illustrate the presence of harmonic structure in sequence~D stars, and to motivate the harmonic-filtered selection used in Section \ref{sec:sample_selection}}
    \label{fig:PL_seqD_D1-2}
\end{figure}

Figure~\ref{fig:PL_seqD_D1-2} shows the period–luminosity (PL) diagram of long-period variables (LPVs) in the Large Magellanic Cloud (LMC), based on the OGLE-III catalog. The periods correspond to the dominant periodicity identified in each light curve, plotted against the Wesenheit index $W_{JK} = K_s - 0.686 (J - K_s)$. Among the known PL ridges, sequence~D is populated by stars exhibiting long secondary periods (LSPs). A secondary ridge is also visible at approximately half the period of sequence~D, which we refer to here as sequence~D1/2. This ridge largely overlaps with the high-amplitude regime of sequence~E, commonly associated with ellipsoidal binary systems \citep{soszynski_2004_ellipsoidal_variability,pawlak_2014_seqE_PL_fit}. However, detailed comparisons of photometric and radial-velocity behaviour have shown that the sequence~D variability itself is not caused by ellipsoidal modulation, and that stars on sequences~D and~E exhibit very distinct observational properties \citep{Nicholls_2010_ellipsoidal_variability_sED}. In the context of this work, the designation D1/2 is therefore used to emphasize the hypothesis that a subset of stars on this harmonic ridge may not be ellipsoidal or eclipsing binaries, but instead reflect harmonic structure arising from the same physical mechanism responsible for the sequence D signal. The joint presence of significant power on both sequence D and its harmonic is a prerequisite for applying the harmonic phase diagnostic developed below, and does not imply that sequence D1/2 represents an independent pulsation sequence or a reinterpretation of sequence E. 

To investigate the harmonic structure of these stars, we reanalyze the OGLE I-band light curves using Lomb–Scargle periodograms. This approach is robust to the irregular OGLE cadence and reveals periodicities and harmonic content. For each star, we normalise the time axis by the primary period (P$_1$) and stack the resulting power spectra. Figure \ref{fig:lomb_scargle_LMC_ogle} shows the results for stars with dominant periods on sequence D (left) and on D1/2 (right). Sequence D stars exhibit a strong second harmonic at half the primary period, with a median amplitude reaching $\sim30\%$ of the fundamental. 
Harmonics in the light curves of sequence D stars have been previously interpreted as strong evidence for binarity \citep{soszyski_2021_binary_origin_LSP}, where secondary minima are attributed to eclipses by an orbiting companion. However, recent work \citep{courtneybarrer_2025_rtpav} demonstrated that oscillatory convective dipole modes, modeled with a $\ell=1$, $m=0$ spherical harmonic temperature profile, including limb-darkening, can produce similar second harmonic in the light curves, but with a secondary maxima rather than minima. This occurs when the dipole axis is edge on relative to the observer so that both hemispheres are partially visible. Instead of the usual minima caused when the distant (non-observable) hemisphere is at a peak, when edge on, it becomes partially visible and causes a secondary maxima in the observed flux. This creates harmonic content at the dipole frequency. Details of the model implementation are outlined in Appendix \ref{ap:dipole_model}, which follow similar workings by \citet{takayama_2020_LSP_thermal_dipole}. 

\subsection{Distinct signatures of binary vs oscillatory convective mode harmonics}\label{sec:phase_differences} 
The fundamental difference between secondary maxima (an inclined oscillatory convective mode), and secondary minima (an eclipsing binary system) can be formally described in terms of the phase difference between the fundamental ($\phi_1$) and its first harmonic. Let a light curve be approximated as:
\begin{equation}
x(t) = A_1 \cos(\omega t + \phi_1) + A_2 \cos(2\omega t + \phi_2),
\end{equation}
and define the relative phase on [$-\pi$,$\pi$] as
\begin{equation} \label{eq:phase_relation}
\Delta\phi = \phi_2 - 2\phi_1.
\end{equation}
The waveform shape is governed by the amplitude ratio $A_2/A_1$ and the relative phase $\Delta\phi$. Appendix \ref{ap:phase_relation_sec_min_v_max} shows qualitatively when considering the light curve flux that:
\begin{align}
\Delta\phi &\approx 0 \;\Rightarrow\; \text{secondary maximum}\\
\Delta\phi &\approx \pm\pi \;\Rightarrow\; \text{secondary minimum}.
\end{align}
This phase difference $\Delta\phi$ provides a robust, model-independent diagnostic: a population of stars showing $\Delta\phi \approx \pm\pi$ is consistent with the binary scenario, while $\Delta\phi \approx 0$ is naturally explained by oscillatory dipole modes viewed at high inclination. $\Delta \phi \approx \pm \pi$ would only be consistent with an oscillatory dipole if there is a more complex non-linear relationship between temperature and brightness than the simple blackbody function used here. In practice, $\Delta\phi$ is stable to OGLE sampling and noise for the harmonic-selected sample (typical uncertainties of order $\leq 10^\circ$). Note that performing this analysis on stellar magnitude, rather than flux reverses this phase condition. All measurements in this work are based on normalised flux light curves (not magnitudes).
\begin{figure}[h]
    \centering
    \includegraphics[width=0.95\linewidth]{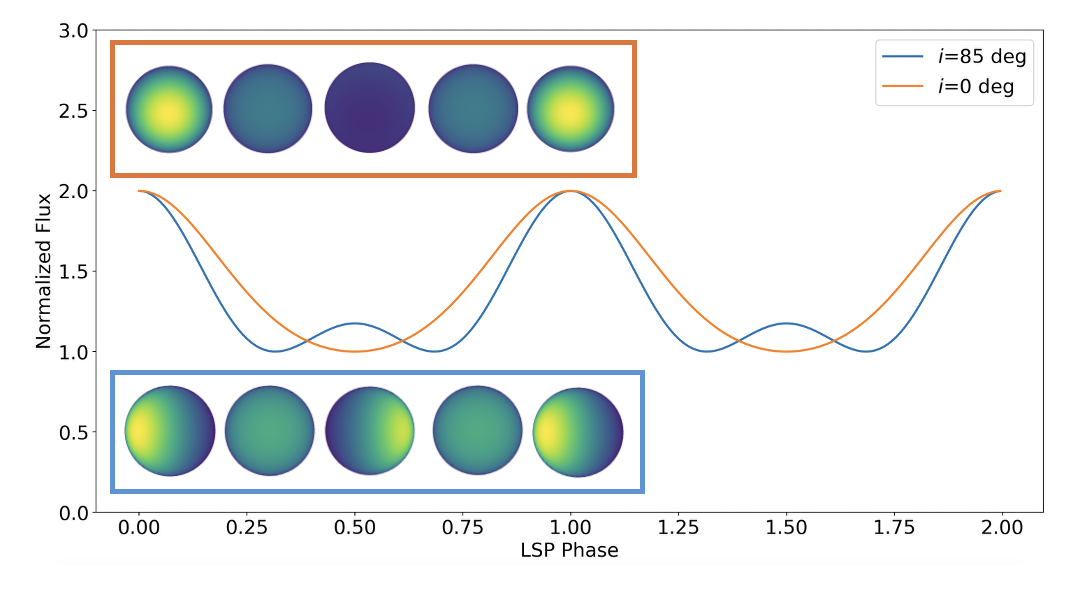}
    \caption{ Example of an inclined vs face-on oscillatory convective dipole causing a secondary maxima in the light curve, therefore producing a harmonic signal.}
    \label{fig:dipole_example_where_harmonic_form}
\end{figure}

\subsection{Harmonic visibility across parameter space}
\begin{figure}[h]
    \centering
    \includegraphics[width=8cm]{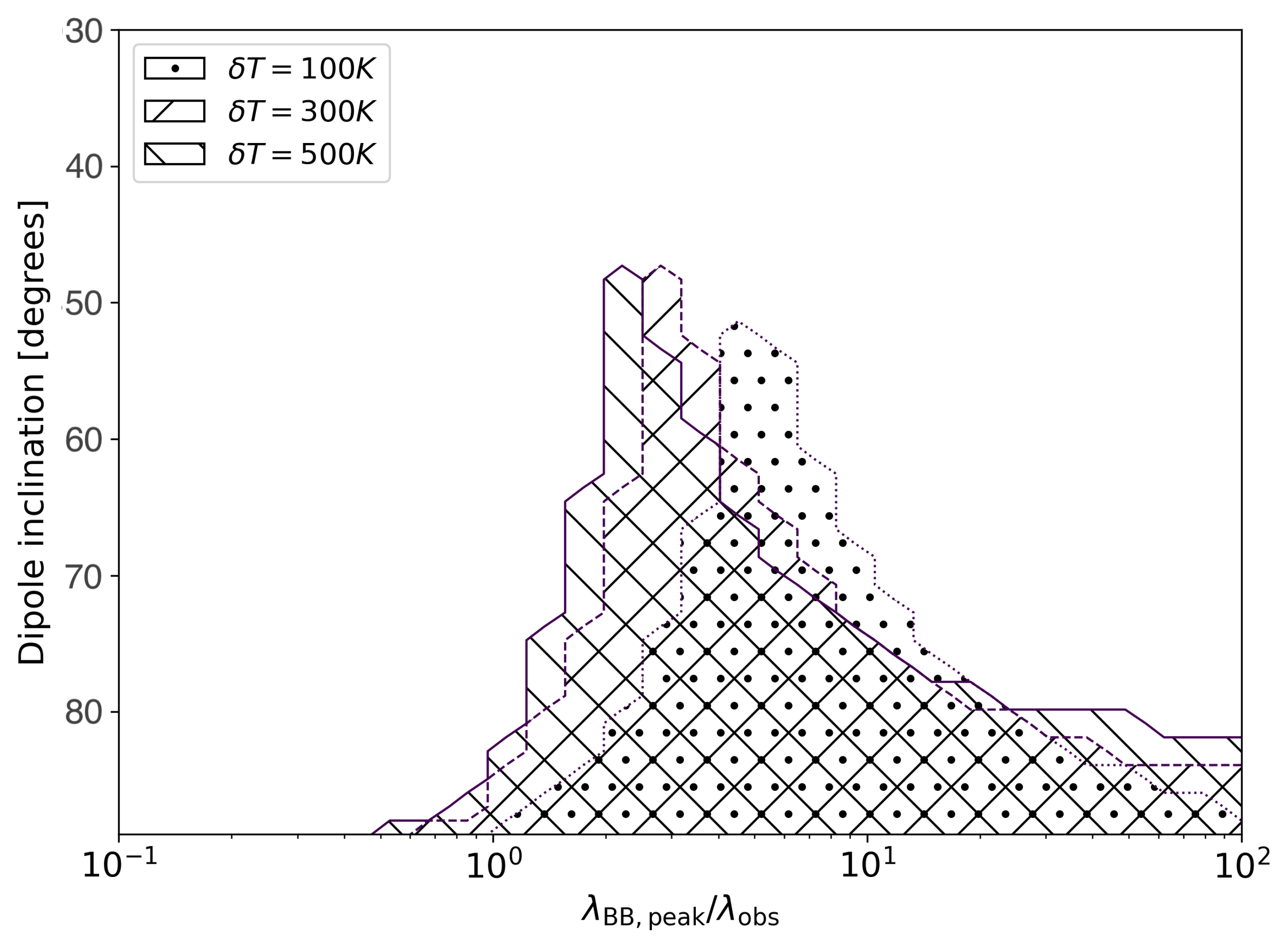}
    \caption{Shaded regions indicate combinations of dipole amplitude, viewing inclination, and $\lambda_{\mathrm{BB,peak}}/\lambda_{\mathrm{obs}}$ where harmonic features appear in synthetic light curves.}
    \label{fig:harmonic_parameter_space}
\end{figure}

To identify when thermal dipole modes produce observable harmonics, we simulate light curves across a grid of different dipole inclination angles $\theta_{\text{obs}}$, temperature amplitudes $\delta T_{\text{eff}}$ and observed wavelength $\lambda_{\mathrm{obs}}$, scaled by the blackbody peak $\lambda_{\mathrm{BB,peak}}$ from Wien's law. A harmonic is deemed “observable” when the secondary peak amplitude exceeds 5\% of the primary, measured directly in the time domain from peak spacings and heights. This approach avoids false positives from sharp asymmetric peaks that contaminate power spectral methods at high $\lambda_{\mathrm{BB,peak}}/\lambda_{\mathrm{obs}}$ ratios. The specific threshold is chosen because in OGLE light curves a 5\% harmonic amplitude typically exceeds the photometric noise floor, even for faint LMC giants. Figure \ref{fig:dipole_example_where_harmonic_form} shows example light curves at high vs. low inclination. Figure \ref{fig:harmonic_parameter_space} maps the parameter space where LSP harmonics are expected to be observable. This observability fundamentally depends on the effective temperature of the star (parameterized as the wavelength $\lambda_{BB,peak}$ of the black body's peak from Wien's law), the observation wavelength $\lambda_{obs}$, the inclination of the dipole and its amplitude $\delta T$. Peak observability is found around $\lambda_{BB,peak}/\lambda_{obs}\sim3$ and no harmonic signature is found for inclinations below 48 degrees. For a typical red giant with $T_{\rm eff} \sim 3500\,\mathrm{K}$, the blackbody peak occurs near $\lambda_{\rm peak} \sim 0.83\,\mu{\rm m}$, implying increased sensitivity to temperature perturbations at shorter visible wavelengths. While this suggests larger fractional variations in the $V$-band compared to $I$, the OGLE $V$-band light curves are more sparsely sampled, limiting their suitability for harmonic phase analysis. The model also predicts phase-dependent colour variations, with larger amplitudes at shorter wavelengths leading to periodic differential colour changes. In addition, the dipole mode is expected to produce radial velocity variations through associated surface flows, including a radial component \citep{saio_2015_oscillatory_convective_modes,takayama_2020_LSP_thermal_dipole}. These provide additional diagnostics for future study.

\subsection{Assumption of no rotation}
In the simulations presented above, we assume the thermal dipole pattern remains fixed in the co-rotating frame of the star and that stellar rotation is negligible over the timescale of the long secondary period (LSP). This implies that the dipole oscillation maintains a constant orientation relative to the observer, modulating the observed flux purely through geometric projection and thermal contrast. This assumption is justified if the stellar rotation period is much longer than the LSP, which has been shown to often be the case in the stellar envelope \citep{olivier_2003_origin_LSP,li_2024_asteroseismic_meas_rotation_rgb}. However, in more rapidly rotating stars, the dipole axis may precess or sweep across the line of sight, modifying the harmonic content and introducing phase shifts in the light curve. We note that the harmonic observability maps shown in Figure \ref{fig:harmonic_parameter_space} reflect only the static, non-rotating case. More discussion on this is presented in Appendix \ref{ap:discussion_on_rotation}. 

%%%%%%%%%%%%%%%%%%%%%%%%%%%%%%%%%%%%%%%%%%%%%%%%%%%

\section{Observational examples}
\subsection{Harmonic filtering and sample selection} \label{sec:sample_selection}
\begin{figure}[h]
    \centering
    \includegraphics[width=8cm]{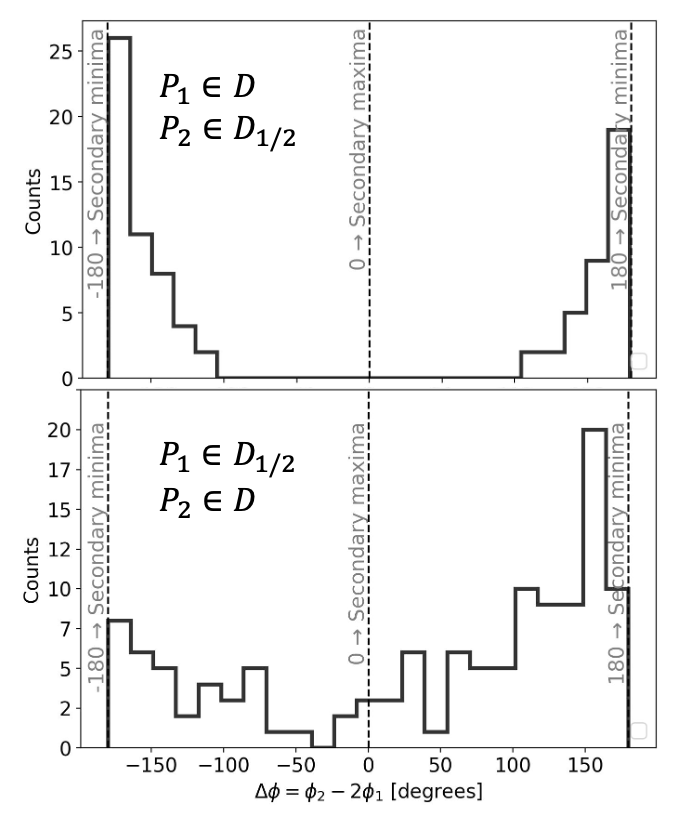}%D-D12_hist.jpeg}
    \caption{Distribution of relative phase differences $\Delta\phi = \phi_{2} - 2\phi_{1}$ for Set 1 ($n=125$, top) and Set 2 ($n=124$, bottom). The peak at $\Delta\phi \approx 0^{\circ}$ indicates secondary maxima (dipole-like), while $\pm180^{\circ}$ indicates secondary minima (binary-like).}
    \label{fig:hist_relative_harmonic_phase}
\end{figure}

We begin with the OGLE catalogue of LSP variables and identify stars whose two strongest periods (P1, P2) correspond to sequence D and the D1/2 ridge. This produces two complementary cases; (1) where the period with the strongest amplitude (P1) is on sequence D, and the period corresponding second strongest amplitude (P2) is at roughly half of P1 (i.e. on sequence D1/2), and (2) where the period with the strongest amplitude is on sequence D1/2, and the period corresponding second strongest amplitude is at twice the period (i.e. on sequence D). We consider a $\sim10\%$ uncertainty on the allowable ratio of $P2/P1$ roughly consistent with the FWHM found in the mean Lomb-Scargle Periodograms (Figure \ref{fig:lomb_scargle_LMC_ogle}). Formally this gives us the following criteria:
\[
\text{(set 1): } P_1 \in D, \quad \frac{P_2}{P_1} \in \left[\frac{1}{2.2},\,\frac{1}{1.8}\right],
\]
\[
\text{(set 2): } P_1 \in D_{1/2},\quad \frac{P_2}{P_1} \in [1.8,\,2.2].
\]
An I-band amplitude threshold (Iamp1 $\ge 0.05$ mag) is also applied to ensure samples have a reasonable SNR.  For each selected star we analyse the OGLE I-band light curve (converted to normalised flux). We fit each OGLE light curve with a multi-sinusoid Fourier model using an iterative frequency-selection procedure. Details are outlined in Appendix \ref{ap:fitting_light_curves}. Histograms of $\Delta\phi$ for each subset are  shown in Fig. \ref{fig:hist_relative_harmonic_phase}.

\subsection{Discussion: Harmonic phase differences and morphology}
Set 1 with the strongest period on sequence D has a distribution of $\Delta \phi$ that strongly peaks around 180 degrees, and no samples $\Delta \phi \sim 0$, indicating known secondary minima consistent with a binary scenario in a circular orbit. However, when considering set 2 (isolating samples with strongest period on sequence D1/2), we find a significantly different distribution that peaks below 180 degrees, and contains $\sim3\%$ of secondary maxima with $\Delta \phi \sim 0$. An example light curve with $\Delta \phi \sim 0$ showing the fitted Fourier model and isolated fundamental and harmonic components is shown in Figure \ref{fig:example_fit}. A second example is shown in Appendix \ref{ap:second_example}, demonstrating that the secondary-maximum morphology is not unique to a single star. These are candidate oscillatory dipoles viewed at high inclination, as predicted by the harmonic content. To verify that secondary maxima in the harmonic-selected sample are physical, we tested the robustness of the relative phase $\Delta\phi$ to photometric noise and the irregular OGLE cadence. Monte Carlo injections of synthetic binary-like light curves ($\Delta\phi=\pi$) demonstrate that noise-driven phase inversions of $\sim180^\circ$ are negligible at the signal-to-noise levels of our sample (Appendix \ref{ap:robustness_deltaPhi}). We also considered contamination from ellipsoidal variables (sequence~E), which overlap the harmonic ridge D1/2. These generally produce near-symmetric light curves with little power in the first harmonic. Such systems are typically filtered out by our harmonic-selection criteria. In the minority of cases where ellipsoidal modulation produces asymmetric light curves with detectable harmonic content, the resulting phase difference remains close to $\Delta\phi\approx\pi$, reinforcing the population of secondary minima rather than mimicking the distinctive secondary maxima ($\Delta\phi\approx0$). Considering a $\sim3000$\,K primary observed in the OGLE I-band ($0.8\,\mu$m), $\lambda_{BB,peak}/\lambda_{obs}\sim1.2$. From Fig.~\ref{fig:harmonic_parameter_space}, the probability that a randomly oriented dipole exceeds the minimum inclination $i_m$ for observability is $P(O)=\cos i_m$. For $\delta T\sim100$\,K \citep{takayama_2020_LSP_thermal_dipole}, we estimate $i_m\approx88^\circ$ ($P\approx0.03$), and for $\delta T=500$\,K, $i_m\approx81^\circ$ ($P\approx0.15$). Using a $\pm10^\circ$ bin around $\Delta\phi=0$, set~2 contains a fraction $\simeq0.03$, broadly consistent with the geometric expectation for $\delta T\sim100$\,K, where harmonics occur only at high inclinations. This comparison is qualitative, as the sample is harmonic-selected and SNR-limited, whereas the inclination estimate applies to the full sequence~D population. The concentration near $\Delta\phi\sim\pm180^\circ$ further indicates that set~2 is a mixed sample with a significant binary contribution. 

We also highlight the concern of \citet{saio_2015_oscillatory_convective_modes} that when adopting a mixing length parameter that is consistent with measured effective temperatures of the red-giant sequences of the LMC, the oscillatory convective modes predicted periods too short by a factor of 2, i.e. exactly where we see the sequence D1/2. It is also here where we observe a broadly consistent fraction of secondary maxima predicted simply by the dipole geometry in inclined systems. However, the use of a single mixing-length parameter to characterise convection throughout the stellar envelope may be overly restrictive.

\begin{figure}[h]
    \centering
%/Users/bencb/Documents/long_secondary_periods/why_galaxies_care_about_agb/fourier_fits_ogle.py
    \includegraphics[width=8cm]{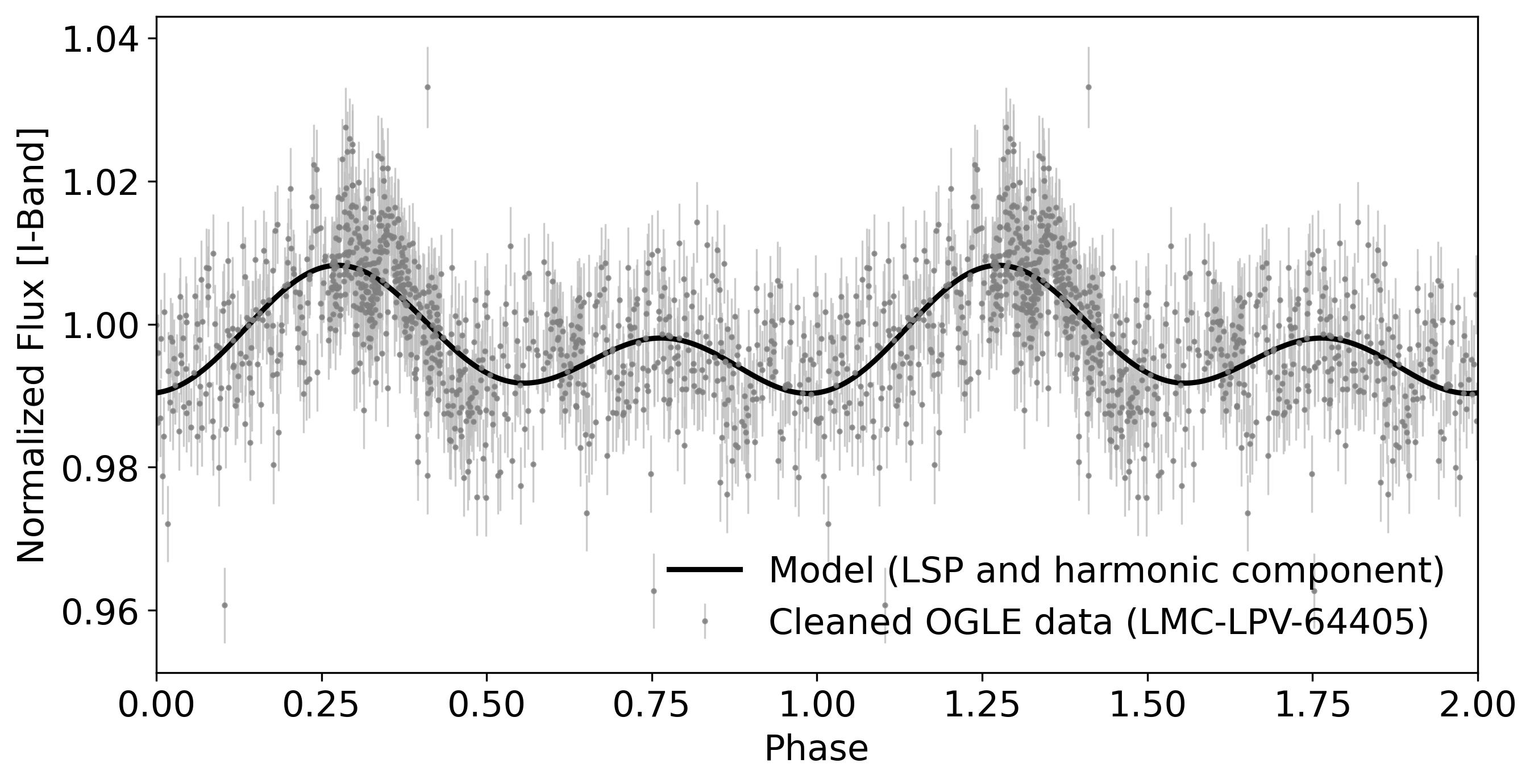}
    \caption{Folded I-band light curve of OGLE-LMC-LPV-64405 filtered for the LSP signal and first harmonic. Grey points show cleaned OGLE data repeated over two cycles. The black curve is the model fit exhibiting a secondary maximum, consistent with an inclined dipole} 
    \label{fig:example_fit}
\end{figure}

%%%%%%%%%%%%%%%%%%%%%%%%%%%%%%%%%%%%%%%%%%%%%%%%%%%
\section{Conclusion}
The relative phase ($\Delta\phi$) of light curve harmonics serves as a powerful diagnostic to distinguish between LSP physical origins. While binary systems produce characteristic secondary minima ($\Delta\phi \approx \pm180^{\circ}$), we show that inclined oscillatory convective dipole modes naturally generate secondary maxima ($\Delta\phi \approx 0$). Our analysis of the OGLE-III LMC catalog reveals that while binarity explains the majority of harmonic-rich LSPs, a statistically non-negligible subset exhibits the secondary-maximum morphology predicted by dipole models. These findings suggest that LSPs represent a mixed population driven by both binary companions and convective oscillations. Future work must now focus on modeling observational selection effects to determine the true relative fraction of each mechanism across the red-giant population.

\begin{acknowledgements}
This work was supported through an Australian Government Research Training Program Scholarship. Special thanks to K.A.E and the SSO.
\end{acknowledgements}

\bibliographystyle{aa} % style aa.bst
\bibliography{bib.bib} % your references Yourfile.bib
\begin{appendix} %First appendix

\section{Lomb-Scargle periodogram of sequence D and D1/2}
\begin{figure}[h!]
    \centering
    \includegraphics[width=8cm]{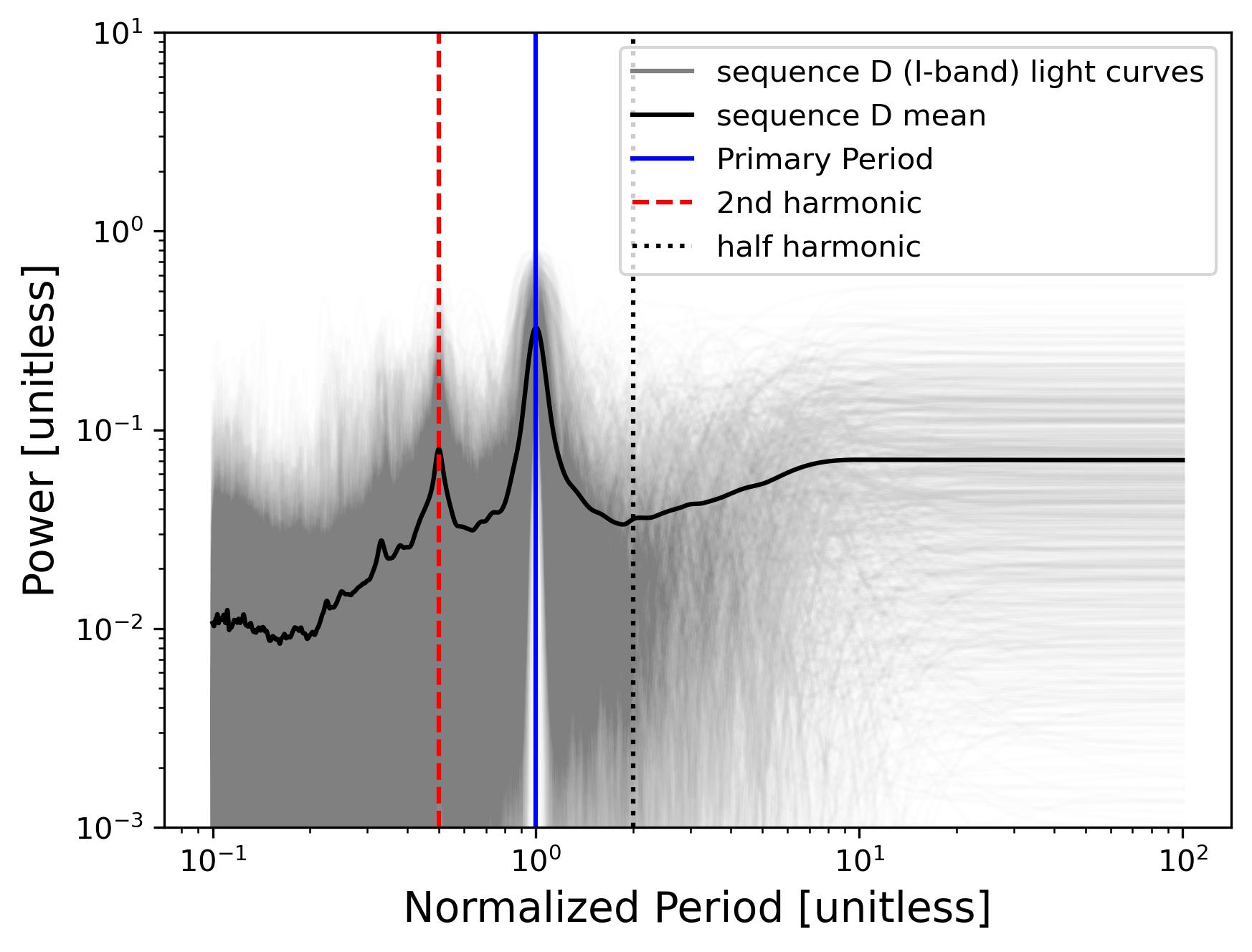}
    \includegraphics[width=8cm]{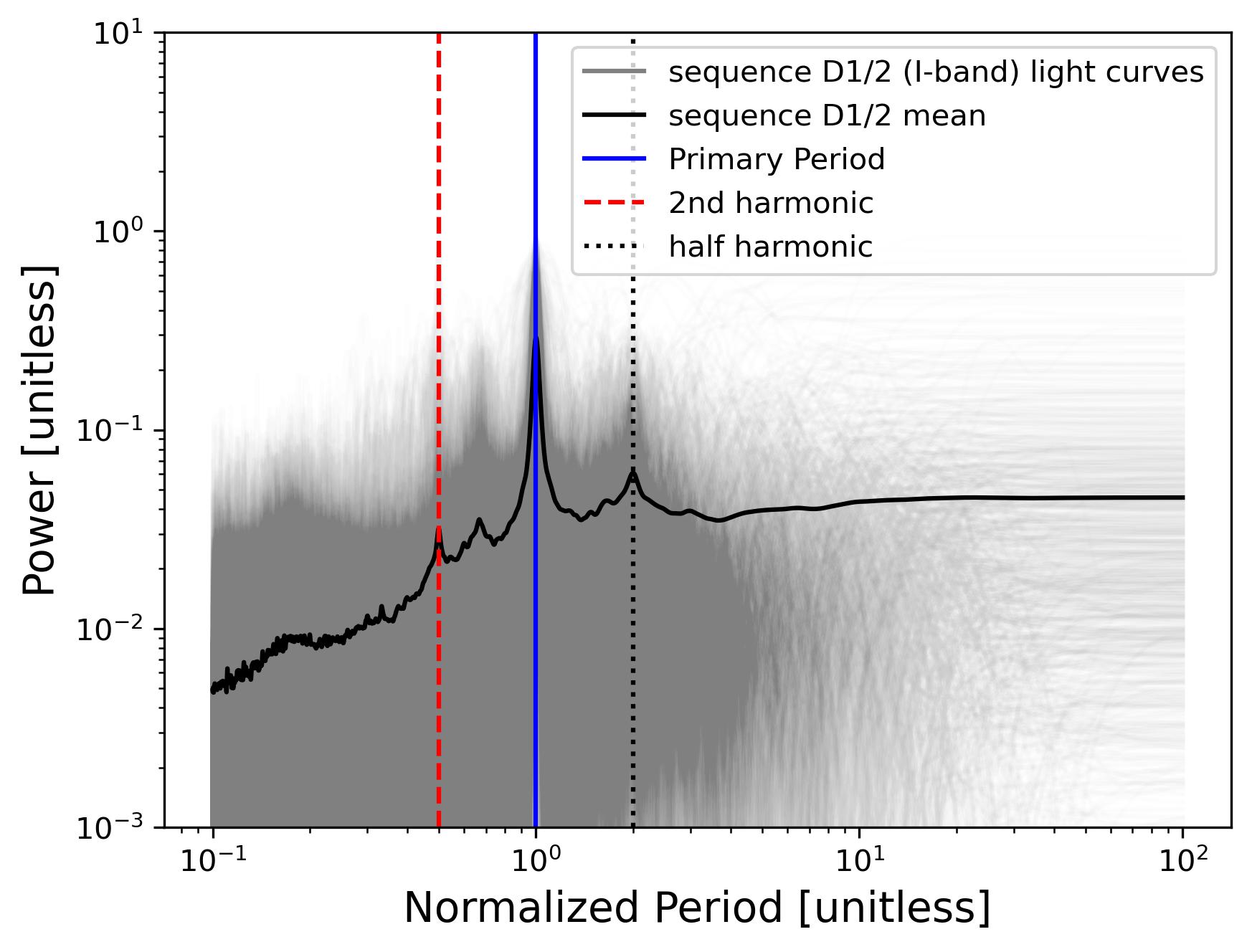}
    \caption{ Lomb-Scargle Periodograms for I-band light curves from the OGLE-III LMC catalog, with time series normalised to the classified primary period, of sequence D (left) and sequence D1/2 (right) LMC stars. }
    \label{fig:lomb_scargle_LMC_ogle}
\end{figure}

Figure \ref{fig:lomb_scargle_LMC_ogle} shows period normalised Lomb-Scargle Periodograms for the sequence D and D1/2 samples.
\section{Oscillatory convective mode modeling}\label{ap:dipole_model}
\subsection{Thermal dipole model}
We model the star’s temperature field as an oscillatory dipole \citep{takayama_2020_LSP_thermal_dipole}, represented by a low-order spherical harmonic:
\begin{align}
    T_{\text{eff,local}} = T_{\text{eff}} + \delta T_{\text{eff}} \, \text{Re}\left[ \frac{Y_l^m(\theta, \phi)}{\max \text{Re}(Y_l^m)} e^{i(2\pi \nu t + \psi_T)} \right],
\end{align}
where $T_{\text{eff}}$ is the mean stellar effective temperature, $\delta T_{\text{eff}}$ is the dipole amplitude, and $\psi_T$ is the phase. The spherical harmonic $Y_l^m(\theta, \phi)$ is normalised to unit peak amplitude and modulates the local surface temperature over time \citep{takayama_2020_LSP_thermal_dipole}.

\subsection{Effective temperature estimates for LMC stars}
To initialize realistic simulations, we estimate effective temperatures for LMC stars using $V$ and $K_s$ photometry from OGLE, corrected for extinction with $A_V = 0.3$ and $A_{K_s} = 0.114 \times A_V$ \citep{gordon_2003_lmc_extinction}:
\begin{align}
    (V - K_s)_0 &= (V - K_s) - (A_V - A_{K_s}).
\end{align}
We then apply the empirical calibration from \citet{Hernandez_2009_giant_Teff_2mass}:
\begin{align}
    \theta_{\mathrm{eff}} &= b_0 + b_1 X + b_2 X^2 + b_3 X\,[\mathrm{Fe/H}] + b_4\,[\mathrm{Fe/H}] + b_5\,[\mathrm{Fe/H}]^2, \\
    T_{\mathrm{eff}} &= \frac{5040}{\theta_{\mathrm{eff}}},
\end{align}
with $X = (V - K_s)_0$ and [Fe/H] = $-0.19$, typical of LMC giants \citep{hocde_2023_metalicity}. 

\subsection{Limb darkening}

To account for disk-integrated flux effects, we include linear limb darkening using coefficients from \citet{claret_2011_limd_darkening}, matched to $T_{\mathrm{eff}}$, $\log g \sim 0$, and [Fe/H] = $-0.4$ typical of AGB stars \citep{cole_2005_metallicity_AGB_LMC}:
\begin{align}
    \frac{I(\mu)}{I(1)} = 1 - u (1 - \mu),
\end{align}
where $\mu = \cos\theta$ is the angle to the line of sight and $u$ is the band-specific limb-darkening coefficient.

\subsection{Projection and observables}

The temperature-modulated, limb-darkened intensity map is rotated into the observer’s frame via:
\begin{align}
    \mathbf{R} = R_z(\phi_{\text{obs}}) R_y(\theta_{\text{obs}}),
\end{align}
followed by projection onto the 2D plane:
\begin{align}
    (x, y) = (\sin \theta_{\text{rot}} \cos \phi_{\text{rot}}, \sin \theta_{\text{rot}} \sin \phi_{\text{rot}}),
\end{align}
with total observed flux computed over the visible hemisphere ($\mu > 0$):
\begin{align}
    F_{\text{obs}}(t) = \int_{\mu > 0} I(\mu, t)\, dA.
\end{align}
The resulting synthetic light curves include geometric modulation, temperature asymmetry, and disk integration.

\begin{figure*}[h!]
    \centering
    \includegraphics[width=0.95\linewidth]{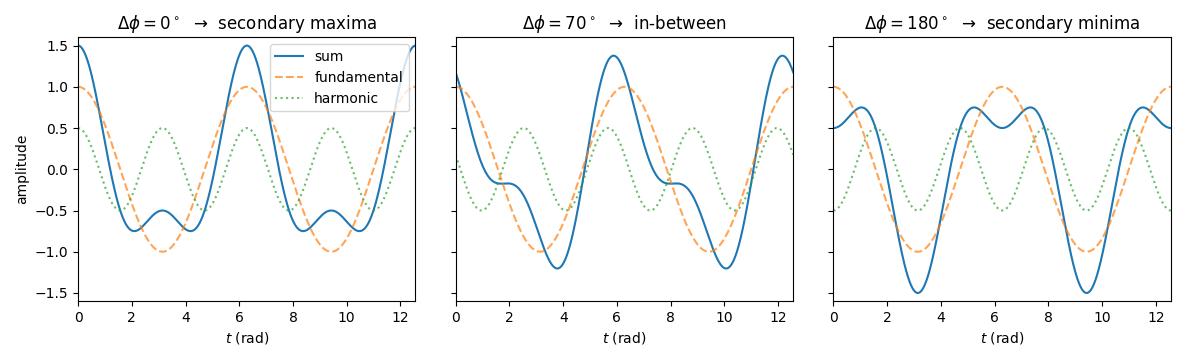}
    \caption{ Example waveforms showing the sum of a fundamental signal and its first harmonic for a fixed amplitude ratio of 0.5. The three panels correspond to relative phase offsets of 0, 60, and 180°, illustrating how changes in the phase relationship between the two components produce distinct waveform shapes—ranging from profiles with secondary maxima to asymmetric peaks and secondary minima.}
    \label{fig:harmonic_phase_relation_secondary_max_v_min}
\end{figure*}

\section{Harmonic phase relation for secondary minima vs maxima}\label{ap:phase_relation_sec_min_v_max}
Beginning with the general waveform:
\begin{equation} \label{eq:harmonic_waveform}
x(t) = A_1 \cos(\omega t) + A_2 \cos(2\omega t + \Delta\phi).
\end{equation}
The waveform shape is governed by the amplitude ratio $A_2/A_1$ and the relative phase $\Delta\phi$. To characterise the nature of the secondary extremum, consider the neighborhood of $t = \tfrac{\pi}{\omega} + \varepsilon$, where the fundamental alone has a minimum. Expanding to second order, the derivatives are:
\begin{align}
x'(t) &= -A_1 \omega \sin(\omega t) - 2A_2 \omega \sin(2\omega t + \Delta\phi), \\
x''(t) &= -A_1 \omega^2 \cos(\omega t) - 4A_2 \omega^2 \cos(2\omega t + \Delta\phi).
\end{align}
Using approximations near $t = \pi/\omega$, we find:
\begin{equation}
\varepsilon^\star = \frac{2A_2 \sin\Delta\phi}{\omega (A_1 - 4A_2 \cos\Delta\phi)},
\end{equation}
The quantity $\varepsilon^\star$ represents the first-order shift in the location of the extremum induced by the harmonic term. Evaluating the curvature at $t = \pi/\omega + \varepsilon^\star$ ensures that the classification as a minimum or maximum is performed at the true stationary point of the combined waveform. The curvature at this stationary point is:
\begin{equation}
x''\!\left(\tfrac{\pi}{\omega} + \varepsilon^\star\right) \approx \omega^2 \left(A_1 - 4A_2 \cos\Delta\phi\right).
\end{equation}
Therefore:
\begin{equation}
\begin{cases}
A_1 - 4A_2 \cos\Delta\phi > 0 \quad \Rightarrow \quad \text{local minimum (secondary dip)}, \\
A_1 - 4A_2 \cos\Delta\phi < 0 \quad \Rightarrow \quad \text{local maximum (secondary peak)}.
\end{cases}
\end{equation}
Evaluating the signal directly at $t = \tfrac{\pi}{\omega}$ gives:
\begin{equation}
x\!\left(\tfrac{\pi}{\omega}\right) = -A_1 + A_2 \cos\Delta\phi.
\end{equation}
Thus, $\cos\Delta\phi > 0$ lifts the minima (tending toward a secondary maximum), while $\cos\Delta\phi < 0$ deepens it (enhancing the minimum). In qualitative terms:
\begin{align}
\Delta\phi \approx 0 \;\Rightarrow\; \text{secondary maximum}\\
\Delta\phi \approx \pm \pi \;\Rightarrow\; \text{secondary minimum}
\end{align}
Higher-order harmonics ($n > 2$) behave analogously, with $\phi_n - n\phi_1$ controlling the location and type of additional extrema. A subtle, but important point is considering when the light curve is in units of magnitude ($X_m$), as in the OGLE catalog, i.e.: 
\begin{equation}
    X_m(t) \propto -\log_{10}(x(t))
\end{equation}
Accordingly the signs are reversed and we get:
\begin{align}
\Delta\phi \approx 0 \;\Rightarrow\; \text{secondary minimum}\\
\Delta\phi \approx \pm \pi \;\Rightarrow\; \text{secondary maximum}
\end{align}

\section{Fourier fitting of OGLE light curves}\label{ap:fitting_light_curves}
After selecting the highest peak (excluding a small neighbourhood around previously selected frequencies), we refit all selected components simultaneously using weighted linear least squares with weights $w_i = 1/\sigma_i^2$, where $\sigma_i$ is the reported photometric uncertainty of datum $y_i$. We solve for an offset and sine-cosine coefficient pairs at each frequency. A newly added frequency is retained only if the Bayesian Information Criterion (BIC) improves, adopting $\mathrm{BIC} = \chi^2 + k \ln N$, where  $\chi^2$ is the standard chi-square statistic, $N$ is the number of data points used in the fit and $k = 1 + 2m$ for $m$ fitted frequencies. The iteration terminates once the BIC no longer decreases. For each accepted component, the sine-cosine coefficients are converted to an amplitude and phase under a cosine convention, allowing evaluation of the phase relation $\Delta\phi$ between a dominant frequency and a near-harmonic counterpart.
\section{Robustness of harmonic phase measurements} \label{ap:robustness_deltaPhi} %/Users/bencb/Documents/long_secondary_periods/why_galaxies_care_about_agb/mc_harmonic_phase_flip_test.py
To assess whether the secondary maxima identified in the harmonic-selected subset of sequence~D stars could arise from noise-induced misclassification, we performed Monte Carlo simulations of synthetic light curves with known harmonic structure. Each simulation consisted of a signal of the form described in \ref{eq:harmonic_waveform} sampled either at the observed OGLE cadence or using a simplified seasonal cadence designed to reproduce the sparse and irregular sampling typical of the survey. Gaussian photometric noise was added at levels matched to the signal-to-noise ratios of the harmonic component in the real sample. For each realization, the fundamental and first harmonic were fit at the known frequencies using a linear least-squares model, and the relative phase $\Delta\phi = \phi_2 - 2\phi_1$ was recovered. This procedure intentionally conditions on the presence of significant power at both the fundamental and harmonic periods, mirroring the harmonic-selection step applied to the OGLE-III data, and isolates the effect of noise and cadence on the recovered phase alone. The simulations therefore do not test the detectability of harmonic power or the reliability of period identification, but specifically address whether noise can cause a true secondary minimum ($\Delta\phi=\pi$) to be misidentified as a secondary maximum ($\Delta\phi=0$).

\begin{figure}[htbp]
%/Users/bencb/Documents/long_secondary_periods/why_galaxies_care_about_agb/mc_harmonic_phase_flip_test.py
    \centering
    \includegraphics[width=0.9\linewidth]{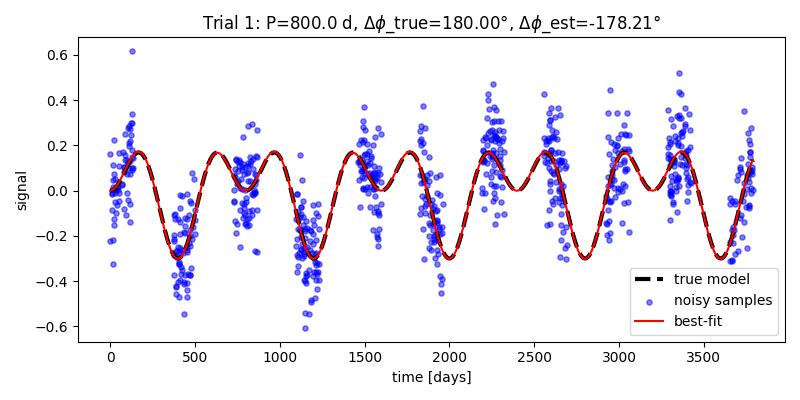}
    \includegraphics[width=0.9\linewidth]{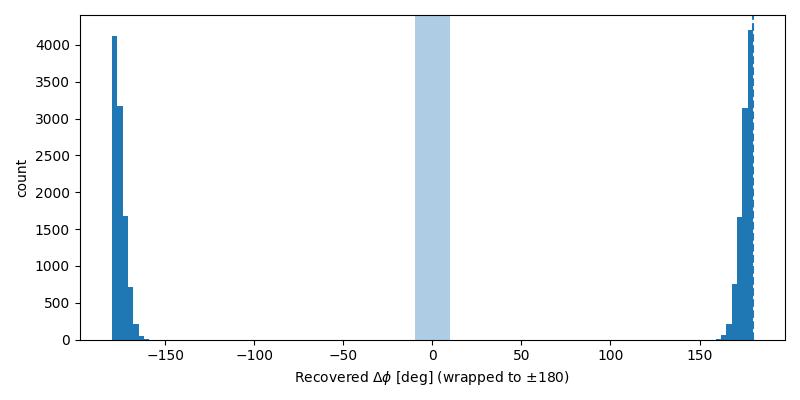}
    \caption{Monte Carlo test of harmonic phase recovery for a synthetic light curve with a true secondary minimum ($\Delta\phi=180^\circ$). 
\textit{Top:} Example realization showing the noiseless model (black dashed), noisy sampled data with seasonal cadence (blue points), and the best-fitting two-harmonic model recovered at the known fundamental and harmonic frequencies (red).
\textit{Bottom:} Distribution of recovered relative phases $\Delta\phi=\phi_2-2\phi_1$ over $2\times10^{4}$ Monte Carlo realizations. The shaded region indicates a $\pm10^\circ$ window around $\Delta\phi=0$, corresponding to the secondary maxima regime. No significant population of noise-driven phase flips from $\Delta\phi=180^\circ$ to $\Delta\phi\approx0$ is observed, demonstrating that such misclassification is negligible at the signal-to-noise levels considered.}
    \label{fig:MC_test_phi_recovery}
\end{figure}

Across $2\times10^{4}$ realizations, the recovered $\Delta\phi$ distribution remains tightly clustered around the true value even with poor cadence (down to tens of days per year), with no statistically significant population of phase flips by $\sim180^\circ$ at the signal-to-noise levels of the observed sample. The recovered phase scatter is well described by a width of order $\sim10^\circ$, supporting the use of a $\pm10^\circ$ bins in the histograms of the main analysis. Example simulated light curves and the corresponding $\Delta\phi$ distributions are shown in Figure \ref{fig:MC_test_phi_recovery}.

\section{Discussion on rotation}\label{ap:discussion_on_rotation}
Our models assumption of non-rotation is generally justified for our red-giant sample, however there are certainly cases where the LSP could be comparable to the rotation period, especially in binary systems. In practice this produces a scatter in $\Delta\phi$, broadening the histogram and occasionally shifting individual stars toward intermediate values. Asteroseismic analysis using \textit{Kepler} data by \citet{ceillier_2017_surface_rotation_redgiants} found that only $\sim$2\% of red giants with solar-like oscillations exhibit detectable surface rotation periods. More recently, \citet{li_2024_asteroseismic_meas_rotation_rgb} conducted a large survey of 2,495 RGB stars and measured core and surface rotation using rotational splittings of mixed modes. Of the detectable periods they reported periods mostly in the 200 - few thousand day range, with core-to-envelope rotation ratios peaking near 20 for RGB stars, indicating significant differential rotation. In this context, the width of our $\Delta\phi$ distribution becomes an important diagnostic: stars near $\Delta\phi \approx 0^\circ$ likely represent dipoles in nearly non‐rotating or slowly rotating envelopes.  Stars with larger deviations may reflect modest rotation-induced phase drifts and/or frequency chirps. Simple geometric models could be included to characterise the harmonic precession  under a dipole hypothesis. There remains an important uncertainty concerning the rotational coupling of the oscillatory convective dipole, and whether this large-scale convective asymmetry is anchored in the more rapidly rotating core, the slowly rotating envelope, or a distinct intermediate shell.

\section{Additional example of secondary maxima} \label{ap:second_example}
\begin{figure}
    \centering
    \includegraphics[width=8cm]{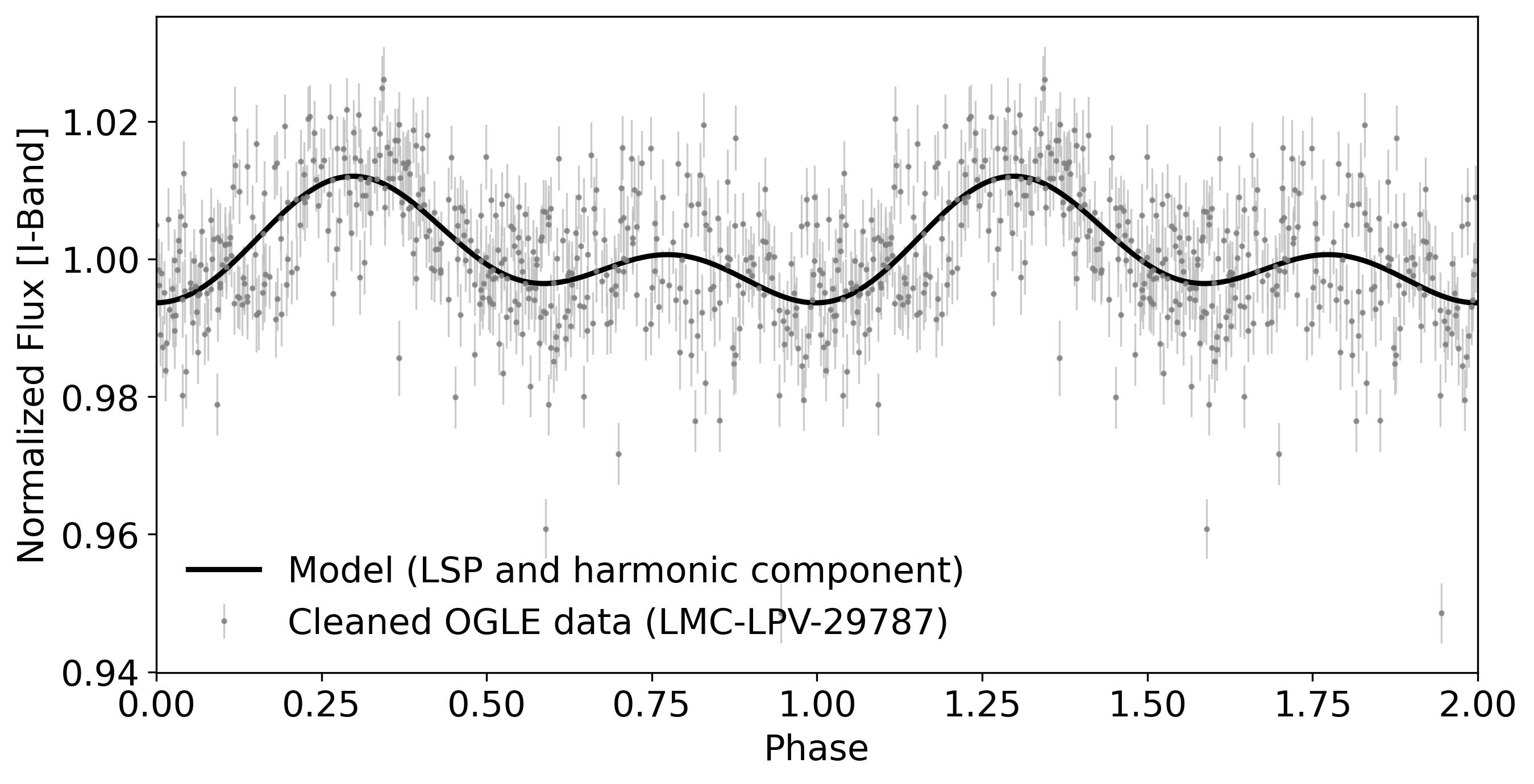}
    \caption{Folded OGLE I-band light curve of OGLE-LMC-LPV-29787. An additional example where the fitted sequence D and harmonic component shows a clear secondary maxima corresponding to $\Delta \phi\sim0$ (dipole mode).}
    \label{fig:2nd_example_fit}
\end{figure}

Figure~\ref{fig:2nd_example_fit} shows a second example from the $\Delta\phi \sim 0$ subset. The folded OGLE I-band light curve exhibits a clear secondary maximum consistent with the inclined oscillatory convective dipole interpretation described in the main text. This demonstrates that the morphology shown in Figure~\ref{fig:example_fit} is not unique to a single object.

\end{appendix}
%-------------------------------------------------------------------
\end{document}